\documentstyle[preprint,aps]{revtex}

\tightenlines 

\title{Controlled dephasing of Andreev states in superconducting
quantum point-contacts}
\author{M. A. Desp\' osito$^*$ and  A. Levy Yeyati$^\dag$}
\address{
* Departamento de F\'{\i}sica, Facultad de
Ciencias Exactas y Naturales,\\ Universidad de Buenos Aires, 1428
Buenos Aires, Argentina. \\
\dag Departamento de F\'\i sica de la Materia Condensada C-V. Facultad de Ciencias, \\
 Universidad Aut\'onoma de Madrid, E-28049 Madrid, Spain.  
}
\begin{document}
\input epsf
\draft
\maketitle

\begin{abstract}

We have studied the relaxation and dephasing processes in a 
superconducting  quantum point contact 
induced by the  interaction  with an electromagnetic environment.
Based on a density matrix approach we obtain the rates for the 
dissipative  dynamics as function of the transmission, the phase difference 
on the contact and the external impedance. 
Our calculation allows to determine the appropriate range of parameters 
for the observation of coherent oscillations in the current through the 
contact.
\end{abstract}


\vspace{.2cm}

In a superconducting quantum-point contact (SQPC) the zero-voltage
transport
properties are determined by the so-called Andreev states within the
superconducting energy gap $\Delta$, which are given by
\cite{FuruBeena}
\begin{equation}
E_{\pm}(\phi) = \pm \Delta \sqrt{1 - \tau \sin^2{(\phi/2)}},
\end{equation}
where $\phi$ is the phase difference between the superconducting
electrodes and $\tau$ is the normal transmission coefficient for
each conduction channel.
Atomic contacts, produced by scanning tunneling microscope or
the mechanically controllable break-junction (MCBJ)
technique, have provided an almost ideal realization of an
SQPC with a few conduction channels whose transmissions can be
determined experimentally \cite{Scheer}. Different methods for
Andreev level spectroscopy have been suggested \cite{Gorelik}.

The Andreev states in a single channel SQPC constitute an interesting
realization of a two-level system with well characterized parameters,
which could be proposed as a solid state realization  
of a quantum qubit \cite{qubits}.
The phase difference through the contact can be fixed by the magnetic 
flux in a
superconducting ring geometry like the one depicted in Fig. 1a.
In addition, by varying the external flux one could prepare the system
in a given linear combination of the Andreev states, in close resemblance
to the case of a spin $1/2$ in a magnetic field.
The phase would be, however, affected by fluctuations originated in the
coupling of the ring with its electromagnetic environment, characterized
by a finite impedance $Z(\omega)$. These fluctuations provide a source
of relaxation and dephasing for the dynamics of our two-level system 
\cite{Lundin}.

The aim of this work is to investigate this dissipative dynamics in order
to determine the conditions for the observation of quantum-coherence
effects in this system.


For a single mode SQPC in the low bias regime
one can neglect the continuous part of the quasi-particle spectrum
and restrict the description to the subgap states, introducing the
following Hamiltonian \cite{Ivanov}
\begin{equation}
\hat{H}_0 = \Delta \left\{ \cos(\frac{\phi}{2}) \, \hat{\sigma}_z
+ r \sin(\frac{\phi}{2}) \, \hat{\sigma}_x \right\}  \,   ,
\label{H0}
\end{equation}
where $\hat{\sigma}_i$ are the Pauli matrix and $r = \sqrt{1-\tau}$
is the normal reflection amplitude of the contact. The Hamiltonian
adopts the form (\ref{H0}) in the basis of ballistic states and becomes
diagonal $\hat{H}_0 = E_{+}(\phi)\, \hat{\sigma}_z$ in the basis of the
Andreev states. The ballistic and the Andreev states are represented in
Fig. 1b.

In the ring geometry, the phase on the contact will be given
by $\phi = \phi_0 - \delta\phi $, where $\phi_0 =(2e/\hbar) \Phi$
(being $\Phi$ the external magnetic flux) and $\delta\phi$ represents
the phase fluctuations due to the electromagnetic environment.
If the impedance viewed by the contact is small compared to 
the resistance quantum 
$R_Q= h/4e^2$ one can expand Eq.(\ref{H0}) to the first order in
$\delta\phi$ to obtain the coupling between the subgap levels
and the environment. 
Then, the Hamiltonian can be written as
$\hat{H}_0(\phi) =\hat{H}_0(\phi_0) + \hat{H}_{c}$, 
where the coupling $\hat{H}_{c} = -\frac{\hbar}{2e}\hat{I}\delta\phi$ 
is proportional to the current operator $\hat{I}$.
Performing a rotation to the basis of the Andreev states, 
the current operator is given by
\begin{equation}
\hat{I} = \frac{e \Delta^2}{\hbar E_{+}(\phi_0)}
\left\{ \sqrt{1-\tau} \, \hat{\sigma}_x
-\tau\sin(\frac{\phi_0}{2})\cos(\frac{\phi_0}{2})\, \hat{\sigma}_z
\right\} \, .
\label{current}
\end{equation}
%


On the other hand, a generic 
environment can be represented by a set of L-C circuits,
with phases $\phi_n$, such that $\delta\phi=\sum_n \phi_n$.
The relevant information of the environment is contained in the
phase-phase correlation function \cite{Ingold}

\begin{equation}
C(t)  =  \sum_{n,m}<\phi_n(t)\phi_m(0)> \\
 =  \frac{1}{R_Q}
\int_{0}^{\infty} d\omega \frac{\Re e Z_t(\omega) }{\omega}
\left\{
\coth (\frac{\beta\hbar\omega}{2}) \cos\omega t - i \sin\omega t
\right\}
\label{corre}
\end{equation}
where $Z_t(\omega)$  is the effective impedance of the circuit as seen
from the contact and $\beta$ is the inverse temperature.


Projection operator techniques  allow to derive a
time-convolutionless generalized master equation (GME) \cite{Chang} for 
the reduced
density matrix $\hat{\rho}$ of the Andreev levels \cite{comment}. Up to second order
in the coupling Hamiltonian this equation is given by
\begin{equation}
\dot{\hat{\rho}}(t) + \frac{i}{\hbar}[\hat{H}_0,\hat{\rho}(t)] =
-\frac{1}{\hbar^2} \int_{0}^{t} d\tau
\left\{
%
%
\left[ \hat{I},\hat{I}(-\tau) \hat{\rho}(t)\right]  C(\tau)
-   \left[ \hat{I},\hat{\rho}(t) \hat{I}(-\tau)\right] C^{*}(\tau)
\right\}     
\label{GME} \,  ,
\end{equation}
where $\hat{I}(\tau)$ denotes the current operator (\ref{current}) in the interaction
representation.

One can extract the evolution equations for the matrix elements of the reduced 
density 
projecting the GME (\ref{GME}) in the $\{|+\rangle,|-\rangle\}$ basis of the Andreev states.
Taking into account that $\rho_{++}+\rho_{--}=1$, and defining $\rho_R$ 
and $\rho_I$ as the  real and imaginary part of the off-diagonal matrix 
element $\rho_{+-}$, we finally get 
the rate equations
\begin{eqnarray}
\label{evolution1}
\dot{\rho}_{+} & = &  -(W^{+} + W^{-}) \rho_{+}  + W^{+} -\Lambda \rho_R  \, ,\\
\dot{\rho}_{R} & = & \Omega \rho_{I} + \eta  \dot{\rho}_{+} \, , \\
\label{evolution2}
\dot{\rho}_{I} & = & -(W^{+} + W^{-}+ \eta\Lambda)  \rho_{I}
- (\Omega + 2\Omega_{r}) \rho_{R} + \eta \Omega_{r}(2\rho_{+}-1) + \Upsilon \, ,
\label{evolution3}
\end{eqnarray}
where $\rho_{+}=\rho_{++}$, 
$\eta= \tau \sin(\phi_0/2)\cos(\phi_0/2)/\sqrt{1-\tau}$ 
and $\Omega=(E_{+}-E_{-})/\hbar$.

The coefficients $W^{\pm}$ entering in Eqs. (\ref{evolution1}) to (\ref{evolution3}) 
are 
the upward and downward transition rates which determine the 
system relaxation. The dephasing, i.e. the 
decay of the off diagonal elements, is also controlled
by the coefficient $\Lambda$, while $\Omega_{r}$ plays the role of a level 
shift as discussed below.

The explicit form of all these coefficients depends on the actual 
impedance $Z_t(\omega)$.
Typically, this impedance would be determined by the measuring circuit,
which could consist of a SQUID inductively coupled to the ring with the
SQPC. We shall denote by $M$ and $L$ the mutual inductance between the ring 
and the SQUID and the self-inductance of the SQUID respectively.
The SQUID is also characterized by a finite resistance $R$. 
The charging effects associated with the contact capacitance can be 
neglected in the low impedance regime that we are considering \cite{Ingold}. 
Under these conditions the real part of $Z_t$ will be given by
\begin{equation}
\Re e Z_t(\omega) = \frac{\omega^2 M^2 R }{R^2 + \omega^2 L^2} \, .
\label{Zt}
\end{equation}

If the environment has a large  cuttoff frequency ($\gamma=R/L \gg  \Delta/\hbar$), 
we can make the Markov approximation in the evolution equations (\ref{evolution1}) to (\ref{evolution3}). 
In this case, the involved coefficients can be written as 
\begin{eqnarray}
\label{coef1}
W^{\mp} & = &  W^{-}_{0}  
\left\{ 
\begin{array}{l}
n(\Omega) + 1    \\
n(\Omega)   \\
\end{array} 
\right.    
\, ,  \\
\Lambda & = &  W^{-}_{0}  \eta \frac{2}{\beta\hbar\Omega}  \,  ,   \\
\Omega_{r} & = & W^{-}_{0} \left\{
\frac{1}{2}\cot(\frac{2}{\beta\hbar\gamma}) -\frac{2\gamma^2}{\beta\hbar}
\sum_{k=1}^{\infty}
\frac{\nu_k}{[\gamma^2-\nu_{k}^2][\Omega^2 + \nu_{k}^2]} \right\} \, , \\
\label{coef2}
\Upsilon &  = &   W^{-}_{0} \eta \frac{\Omega}{\gamma} \, ;
\label{coef3}
\end{eqnarray}
where 
\begin{equation}
W^{-}_{0} = \frac{\pi}{4}\frac{\Delta}{\hbar} \frac{\Re e Z_t(\Omega)}{R_Q} 
\frac{1-\tau}{(1 - \tau \sin^2{(\phi/2)})^{3/2}}  \, 
\end{equation}
is the downward transition rate at zero temperature, $n(\Omega)$ is the Bose mean 
occupation number and $\nu_{k}=2\pi k/\beta\hbar$ are the Matsubara frequencies.

The actual value of the coefficients is controlled by the environment
parameters $M$, $L$ and $R$. The experimental set up can be designed in order
to fix these parameters in the appropriate range to observe 
quantum-coherence effects. Longer decoherence times would be obtained by 
reducing $M$ as much as possible. However, SQUID parameters must satisfy
certain constraints, imposed by the need to avoid hysteresis and
thermal fluctuations \cite{Tinkham}. These considerations permit to
estimate a minimal value of $M$ of the order of $0.5 nH$,
with $L= 0.1 nH$ and $R = 20 \Omega$ \cite{Urbina}.
Assuming that the superconducting material is aluminum
($\Delta_{Al} \sim 0.18 meV$), one obtains values of 
$W^{-}_{0}$ in the $n$sec$^{-1}$ range.This rate
fixes the order of magnitude of the coefficients (\ref{coef1}) to (\ref{coef3}).


The system of differential equations 
(\ref{evolution1}-\ref{evolution3}) is characterized
by a set of eigenvalues which determine the typical times for the dissipative
dynamics. In the limit of small impedance these eigenvalues
are given by the following simple expressions
\begin{equation}
\lambda_1 =  - (W^+ + W^-) \, , \quad    
\lambda_{2,3} = -\frac{1}{2}(W^{+} + W^{-} + 2\eta\Lambda) 
\pm i (\Omega + \Omega_{r}) .
\label{eigenvalues}
\end{equation}


Notice that due to the presence of the environment, the Andreev levels 
are shifted according to
$\tilde{E}_{\pm} \simeq E_{\pm} \pm \Omega_{r}/2$.
However, for the range of parameters we are considering this renormalization
is small  and can be neglected.

Solving the equations (\ref{evolution1}-\ref{evolution3}) at the lowest order in 
$\Omega^{-1}$ (which is equivalent to implement 
the rotating wave approximation \cite{RWA})
one realizes that the eigenvalues  (\ref{eigenvalues}) are directly 
connected to the relaxation and dephasing rates by the relations
$\Gamma_R \simeq -\lambda_1$ and $\Gamma_D \simeq
-\Re e \{{\lambda}_{2,3}\}$.

The approximate evolution equations can then be written as
\begin{eqnarray}
{\rho}_{+} & = & \frac{W^{+}}{\Gamma_R}(1-e^{-\Gamma_R t})+  \rho_{+}(0) e^{-\Gamma_R t}    \, ,   \\
{\rho}_{R} & = & e^{-\Gamma_D t} ( \rho_{R}(0)\cos\Omega t +  \rho_{I}(0)\sin\Omega t ) \, ,        \\
{\rho}_{I} & = & e^{-\Gamma_D t}  
( \rho_{I}(0)\cos\Omega t - \rho_{R}(0)\sin\Omega t )  \, .
\label{apevol}
\end{eqnarray}
which clearly shows that the diagonal elements of the density matrix 
decay exponentially towards their equilibrium values on a time scale given
by $1/\Gamma_R$ while the non-diagonal elements perform oscillations with
frequency $\Omega$ which are damped on a typical time $1/\Gamma_D$.

This behavior is reflected in the evolution of the current mean value 
which is given by
\begin{eqnarray}
\langle I(t)\rangle & = & I_0 (1-2\frac{W^{+}}{\Gamma_R}) 
+ 2 I_0 e^{-\Gamma_R t} (\frac{W^{+}}{\Gamma_R}-\rho_{+}(0))  
\nonumber\\
&& + 4e \left(\frac{\Delta}{\hbar}\right)^2 \frac{\sqrt{1-\tau}}{\Omega}  
e^{-\Gamma_D t} ( \rho_{R}(0)\cos\Omega t +  \rho_{I}(0)\sin\Omega t ) \, ,
\label{evocurrent}
\end{eqnarray}
where $I_0 = e(\Delta/\hbar)^2 \tau \sin{\phi}/\Omega$ is the equilibrium 
current at zero temperature.

According to Eq. (\ref{evocurrent}) quantum coherence would 
manifest as damped Rabi oscillations in the mean current through the contact.
These oscillations appear for an initial condition in which both the lower and
the upper Andreev state are populated. An arbitrary initial
condition can be in principle reached from the equilibrium situation by 
imposing a given time-dependent evolution to the phase. 
The amplitude of the oscillations are maximal for $\phi=\pi$ where
they become transmission independent. However, by varying the transmission, the
frequency of the oscillations could be tuned in order to reach an
experimentally accessible range. For instance, in the case of aluminum
and transmission $\tau = 0.99$ the frequency $\Omega$ would be of the
order of $3 \times 10^{10} Hz$.

The decay of these oscillations is controlled by the dephasing rate.
Fig. 2 illustrates the dependence of the relaxation and dephasing rates
on the phase $\phi_0$, the contact transmission $\tau$ and the temperature.
At zero temperature we have $\Gamma_R = 2\Gamma_D =W^{-}_{0}$, which
means that both rates reach a maximum at $\phi_0 = \pi$ regardless of
the contact transmission. This maximum decrease with transmission
as $\sqrt{1-\tau}$.
At finite but low temperatures we have
$\Gamma_R = W^{-}_{0} (1+2e^{-\hbar\Omega/kT})$
and  $\Gamma_D =W^{-}_{0}(\frac{1}{2} + 2\eta^2 k T/\hbar\Omega)$.
As expected both rates increase with temperature but the dephasing rate
increases linearly instead of exponentially. 
The maximum of $\Gamma_R$ is located at $\phi_0=\pi$ for all
range of temperatures and behave as
$\sqrt{1-\tau}\coth{\beta\Delta\sqrt{1-\tau}}$. On the other hand,
$\Gamma_D$ develops a dip at $\phi_0 = \pi$ and exhibits a double peaked
structure located
$\phi_0 = 2\arccos{\pm \sqrt{1+(\sqrt{(1-\tau)/2}-1 )/\tau}}$
for high temperatures. In this classical limit we get
$\Gamma_R = W^{-}_{0} 2 k T/\hbar\Omega$ and
$\Gamma_D =\Gamma_R (\frac{1}{2}+ \eta^2)$.


In summary, we have studied the dissipative dynamics of the Andreev states in
a SQPC coupled to an electromagnetic environment characterized by a generic
frequency-dependent impedance. Our results show that with the appropriate choice
of the environment parameters one can reach values of the dephasing and relaxation
rates within the $n$sec$^{-1}$ range. 
A SQPC in a ring geometry thus provides an interesting realization of a
solid-state two-level system in which
quantum interence effects could be observed.

We acknowledge useful discussions with C. Urbina, D. Esteve and A. 
Mart\'{\i}n-Rodero.
We are also indebted to F. Vazquez and A. Raguet for their stimulating
comments.
This work was supported by the Spanish CICyT under contract PB97-0044.


\begin{figure}
\vspace{5cm}
\begin{center}
\leavevmode
\epsfysize=8cm
\epsfbox{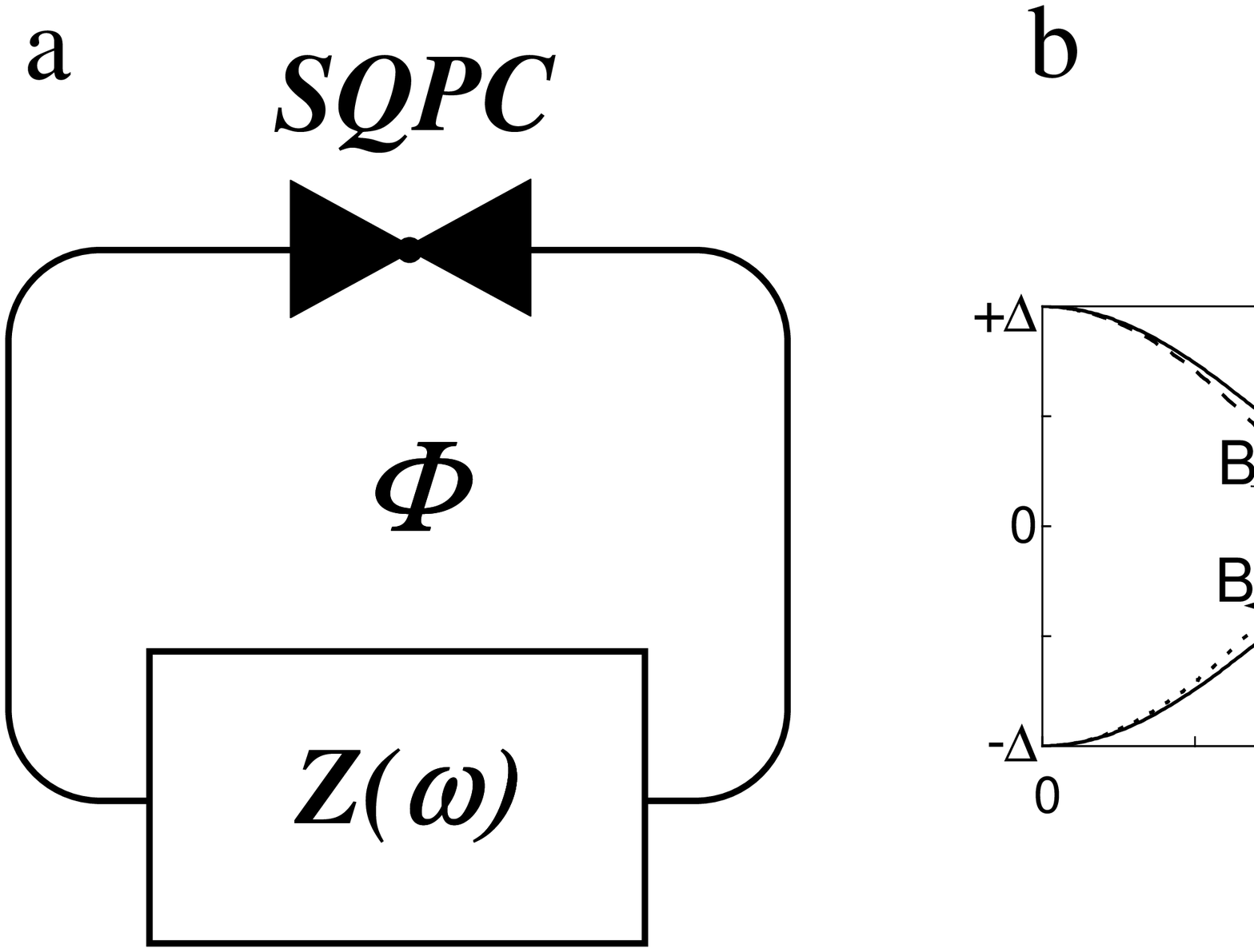}
\end{center}
\caption{
(a) Schematic circuit: a superconducting quantum point-contact (SQPC) inserted in a
superconducting ring threaded by a magnetic flux $\Phi$. $Z(\omega)$ denotes the
effective impedance seen by the contact. 
(b) Andreev levels with phase dependent energies $E_{\pm}$ (full lines).
The dotted and dashed lines indicate the ballistic states.}
\end{figure}

\newpage

\begin{figure}
\centering
\epsfxsize=\columnwidth
\epsfysize=18cm
\epsfbox{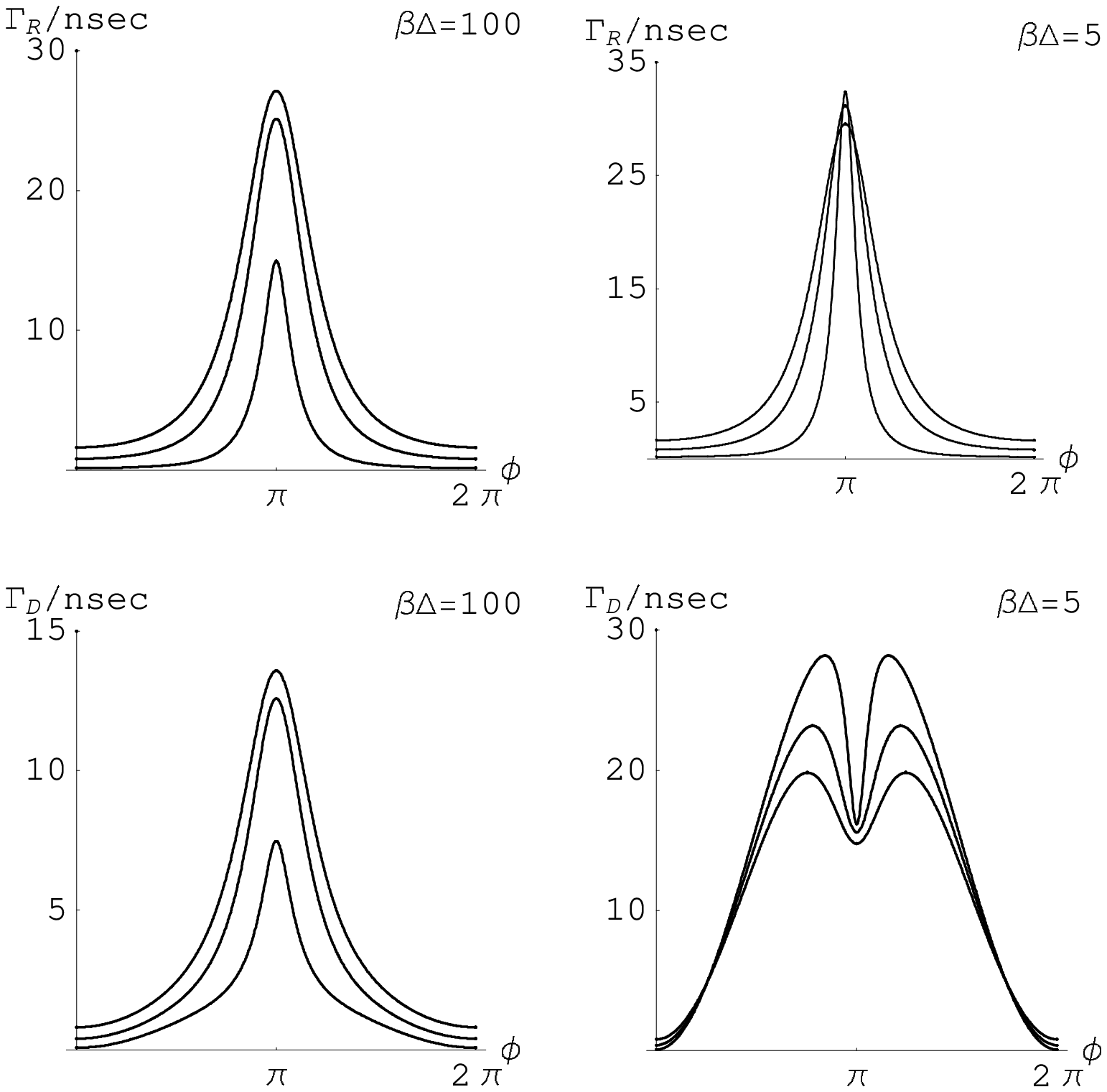}
\caption{Relaxation and dephasing rates $\Gamma_R$ and $\Gamma_D$ as function of the phase $\phi$
for transmission coefficients $\tau$= 0.9, 0.95 and 0.99 (from top to bottom)
and two different values of the parameter $\beta\Delta$.
The downward transition rate at zero temperature $W^{-}_{0}$ 
coincides in the graphic scale with $\Gamma_R$ for $\beta\Delta=100$.}
\end{figure}


\begin{references}
\bibitem{comment}
Notice that our analysis is restricted to the subgap states
neglecting coupling with the continuous
part of the spectrum. Other interactions, such a electron-phonon
coupling, would  introduced a broadening of the subgap states \cite{subgap}.
For aluminum atomic contacts this broadening is, however, small 
\cite{Goffman} and would not introduce important corrections to the 
relaxation and dephasing rates.
\end{references}
\end{document}